\documentclass[aps,twocolumn,showpacs,amsmath,amssymb,pre,superscriptaddress, floatfix,longbibliography]{revtex4-1}

\usepackage{graphicx}% Include figure files
\usepackage{dcolumn}% Align table columns on decimal point
\usepackage{bm}% bold math
\usepackage{amssymb}
\usepackage{multirow}

\begin{document}

\title{Chaotic and ballistic dynamics in time-driven quasiperiodic lattices.}

\author{Thomas Wulf}
    \email{Thomas.Wulf@physnet.uni-hamburg.de}
    \affiliation{Zentrum f\"ur Optische Quantentechnologien, Universit\"at Hamburg, Luruper Chaussee 149, 22761 Hamburg, Germany}

\author{Peter Schmelcher}
    \email{Peter.Schmelcher@physnet.uni-hamburg.de}
    \affiliation{Zentrum f\"ur Optische Quantentechnologien, Universit\"at Hamburg, Luruper Chaussee 149, 22761 Hamburg, Germany}
    \affiliation{The Hamburg Centre for Ultrafast Imaging, Universit\"at Hamburg, Luruper Chaussee 149, 22761 Hamburg, Germany} 

\date{\today}

\pacs{05.45.Ac, 05.45.Pq, 05.45.Gg}

\begin{abstract}

We investigate the nonequilibrium dynamics of classical particles in a driven quasiperiodic lattice based on the Fibonacci sequence. 
An intricate transient dynamics of extraordinarily long ballistic flights at distinct velocities is found. We argue how these transients are caused and can be understood by a hierarchy 
of block decompositions of the quasiperiodic lattice.
A comparison to the cases of periodic and fully randomized lattices is performed.

% Thereby, we demonstrate how each level of the hierachial decompositions introduces a highly nontrivial lengthscale 
% we introduce a suitable set of Poincare surface of sections by whom 

\end{abstract}

\maketitle
%%%%%%%%%%%%%%%%%%%%%%%%%%%%%%%%%%%%%%%%%%%%%%%
\paragraph*{Introduction.}
One of the workhorses in the field of classical chaotic dynamics of time-driven setups are driven lattices, i.e. spatially periodic potentials subjected to 
a temporally periodic forcing. In particular, their resulting transport properties have received tremendous attention as these provide working principles for Brownian or molecular motors 
\cite{hanggi_artificial_2009, denisov_tunable_2014, serreli_molecular_2007, matthias_asymmetric_2003, reimann_brownian_2002, astumian_fluctuation_1994, cubero_hidden_2016, schmitt_molecular_2015, astumian_thermodynamics_1997}
or have even found direct technical applications, e.g. in particle species separation \cite{savelev_separating_2005, bogunovic_particle_2012, roth_optimization_2015, tahir_dynamically-tunable_2014}. 
In terms of the direct experimental realization of such driven lattice potentials, cold atoms 
loaded into shaken optical lattices as generated by counter propagating laser beams have been shown to provide an ideal toolbox as they allow for precise control of the system parameters and thus 
for an experimental verification of many of the theoretically introduced concepts on ratchet transport \cite{sadgrove_engineering_2013, white_experimental_2013, gommers_quasiperiodically_2006, wickenbrock_vibrational_2012}.

An interesting aspect of driven lattice physics has been the effect of deviations from a purely periodic setup. Here, for coupled, dissipative systems it was shown how isolated impurities may stabilize
soliton solutions \cite{alexeeva_impurity-induced_2000, yu_resonant_2011} or how -more generally- the introduction of disorder initiates synchronization in the asymptotically 
reached state \cite{braiman_taming_1995, lei_disorder_2012}.
Recently, it was also demonstrated how disorder may lead to
ordering, in the sense of increased autocorrelations and strongly peaked velocity distributions, even in Hamiltonian lattice systems \cite{wulf_disorder_2014}.
At this point it is certainly worth mentioning that the two structurally limiting cases of strictly periodic and fully randomized lattices have, of course, also been investigated keenly in the quantum domain. Here
the periodic regime is characterized by extended Bloch waves \cite{bloch_uber_1929}, whereas randomness is often accompanied by the celebrated Anderson localization effect \cite{anderson_local_1978}. 
In the quantum domain however, a third form of spatial structure 
has also attracted considerable attention, namely quasiperiodic lattices, triggered particularly by the pioneering work of Shechtman et al. \cite{shechtman_metallic_1984} where
the possibility of long range order even in the absence of translational symmetry was demonstrated. 
In fact, it was shown how quasiperiodicity leads here to a qualitatively new phenomenology compared to 
both the periodic and the random cases \cite{iguchi_theory_1991, kohmoto_localization_1983}, specific examples being self similar critical states or singularly continuous energy spectra \cite{kohmoto_critical_1987, maci&xe1_nature_2014}. 

In the classical regime however, quasiperiodic lattices and their associated chaotic nonequilibrium dynamics have 
so far been largely unexplored. Shining light on this dynamics is the purpose of the present manuscript.
To this end, we study periodically driven lattices build of individual scatterers which are arranged in a quasiperiodic, 
as compared to a periodic or randomized, manner. 
We hereby focus on
quasiperiodicity as generated by the Fibonacci sequence, constituting one of the commonly studied implementations of quasiperiodicity \cite{maci&xe1_nature_2014}. 
Indeed, we showcase observables, in particular the ballistic flight length distribution, where qualitative differences between the quasiperiodic and the periodic and random lattices are apparent. 
Specifically, we find velocity domains where particles perform exceptionally long ballistic flights, a feature shown to be absent in the randomized lattice and hence hinting at the high degree of 
long range order in the Fibonacci chain \cite{zhi-xiong_properties_1994, droubay_palindromes_1995, morfonios_local_2014}. 
We demonstrate how the quasiperiodic lattice can be decomposed into a hierarchy of building blocks, where each hierarchy is shown to naturally induce 
a set of Poincar\'e maps which describe the dynamics on increasingly larger length scales. By this approach, we are able to relate invariant subsets of the Poincar\'e maps 
corresponding to distinct hierarchical levels 
to the observed long ballistic flight events. 
Here we stress that the routinely employed analysis tools, in particular Poincar\'e surfaces of section, rely intrinsically on the driven systems periodicity. Hence, they cannot be applied straightforwardly
to driven quasicrystalline systems, making their analysis and physical interpretation of obtained results a genuinely challenging prospect. 
For this reason, we believe that the introduced notion of a set of Poincar\'e maps, adapted specifically to the given quasiperiodic lattice, 
should be of conceptual interest in the investigation of the chaotic dynamics of aperiodic driven systems.  

Our manuscript is structured as follows: In Sec. \ref{sec1} we introduce the driven lattice Hamiltonian for the periodic, quasiperiodic and randomized cases. In Sec. \ref{sec2}, some 
basic notions of chaotic dynamics in driven lattices are introduced. Sec. \ref{sec3} contains a comparison of the flight length distributions for all three cases. These results are further analyzed 
in Sec. \ref{sec4} and explained by means of a block decomposition of the Fibonacci lattice in Sec. \ref{sec5}. Finally, we provide our conclusions in Sec. \ref{sec6}.

% Thereby, one of the cornerstones of the theoretical analysis is often the Poincare surface of section which are essentially cuts 

%%%%%%%%%%%%%%%%%%%%%%%%%%%%%%%%%%%%%%%%%%%%%%%%%%%%%%%%%%%%%%%%%%%%%%%%%%%%%%%%%%%%%%%%%%%%%%%%%%%%%%%%%%%%%%%%%%%%%%%%%%%%%%%%%%%%%%%%%%%%%%%%%%%%%%%%%%%%%%%%%%%%%%%%%%%%%%%%%%%%%%%%%%%%%%%%%
\section{The driven lattice Hamiltonian.}
\label{sec1}

Throughout this work we study the dynamics of noninteracting classical particles of equal mass $m$ governed by the driven lattice Hamiltonian:

\begin{equation}
  H(x,p,t)= \frac{p^2}{2m}+ \sum_{n=1}^{\infty} V_n \cdot \Theta (l/2-|x-X_{n}-d(t)|).
  \label{Hamiltonian}
\end{equation}

\begin{figure}[t!]
\centering
\includegraphics[width=1\columnwidth]{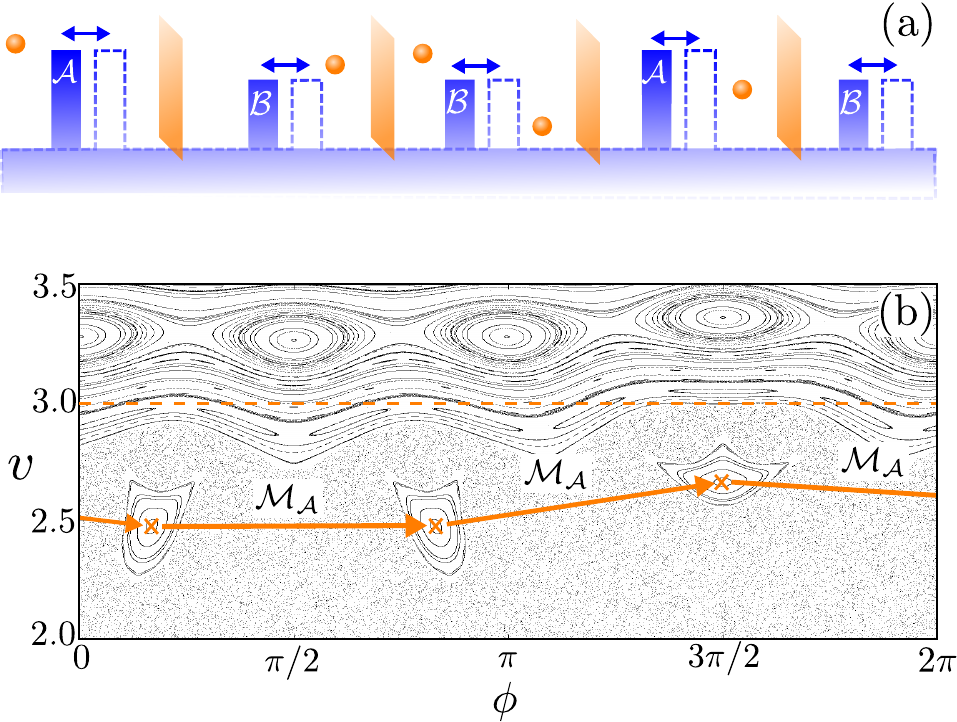}
\caption{\label{fig1} (a) Sketch of a laterally oscillating lattice build of two barrier types $\mathcal{A}$ and $\mathcal{B}$ distinguished by their potential heights $V_\mathcal{A}$ and $V_\mathcal{B}$.
Between consecutive barriers are the Poincar\'e surfaces (orange).
(b) Extract of the Poincar\'e surface of section of a periodic lattice consisting only of $\mathcal{A}$-type barriers with $V_\mathcal{A}=1.5$. The horizontal dashed line indicates the maximal velocity $v^{\text{C}}_{\text{max}}$
that particles can acquire in the chaotic sea. Also shown is the period three fixed point of the Poincar\'e map $\mathcal{M}_\mathcal{A}$ centering the three corresponding regular islands.
Each arrow indicates one application of $\mathcal{M}_\mathcal{A}$ on a trajectory in the fixed point.
Remaining parameters are: $\omega=A=m=l=1.0$ and $L=5.0$.} 
\end{figure}

That is, the potential consists of an semi-infinite array of individual barriers of width $l$ and site-dependent heights $V_n$.
Furthermore, the barriers oscillate around their equilibrium positions $X_n\equiv n\times L$, where $L$ denotes the lattice spacing, according to the driving law $d(t)=A\sin(\omega t)$
with driving amplitude $A$, driving frequency $\omega$ and resulting temporal periodicity $T=2\pi/\omega$.
(Throughout the manuscript, initial conditions will be chosen randomly at large $x$, such that the boarders of the lattice are not reached within the simulation time).
Such a Hamiltonian may be seen as minimalistic model for time-dependent lattice systems as occurring in radiated semiconductors or in cold atom physics. The major advantage of it being, that 
via appropriate choices of the site dependent barrier heights $V_n$, different spatial structures of the lattice can be realised and dynamical processes occurring in these can be analysed and compared. 
We are here interested in three different types of lattices which will be shown to yield substantially different dynamical evolutions for particle ensembles.
Specifically, these three cases are: periodic lattices (PL), randomized lattices (RL)
and quasiperiodic Fibonacci lattices (FL). Each of those can be realised by introducing two types of barriers denoted symbolically by $\mathcal{A}$ and $\mathcal{B}$. Barriers of different type 
are thereby distinguished by their height, i.e $V_n$ takes either of the two different values $V_\mathcal{A}$ or $V_\mathcal{B}$ throughout the lattice (see Fig.\ref{fig1} (a) for a sketch of the setup). 
$\mathcal{A}$- and $\mathcal{B}$-barriers are then arranged in a periodic, quasiperiodic or randomized manner:
\begin{eqnarray}
\text{PL: } V_n &=& V_\mathcal{A} \nonumber \\
\text{RL: } V_n &=& V_\mathcal{A},\mbox{ for }\sigma_n =1,\quad  V_n=V_\mathcal{B} \mbox{ for }\sigma_n =0\\
\text{FL: } V_n &=& V_\mathcal{A},\mbox{ for }\mathcal{F}_n =1,\quad  V_n=V_\mathcal{B} \mbox{ for }\mathcal{F}_n =0 \nonumber 
\label{Potential}
\end{eqnarray}
% \begin{equation}
% \begin{aligned}
% \text{PL: } V_n &=& V_\mathcal{A}\\
% \text{RL: } V_n &=& V_\mathcal{A},\mbox{ for }\sigma_n =1,\quad  V_n=V_\mathcal{B} \mbox{ for }\sigma_n =0\\
% \text{QL: } V_n &=& V_\mathcal{A},\mbox{ for }\mathcal{F}_n =1,\quad  V_n=V_\mathcal{B} \mbox{ for }\mathcal{F}_n =0
% \label{Potential}
% \end{aligned}
% \end{equation}

where $\sigma$ is a randomized sequence of zeros and ones. Contrarily, $\mathcal{F}$ is a quasiperiodic sequence, again of zeros and ones but whose elements $\mathcal{F}_n$
are arranged according to a construction principle based on the Fibonacci numbers (see e.g. \cite{morfonios_local_2014} for details), such that the first few elements are given by: 
\begin{equation}
 \mathcal{F}= 1\ 10\ 101\ 10110\ 10110101...
 \label{sequence}
\end{equation}
Interestingly, the Fibonacci sequence, although never periodically repeating, 
contains a plethora of structurally highly nontrivial properties, such as local parity symmetry on all scales \cite{morfonios_local_2014}, 
and has been the subject of intensive research in both physics \cite{maci&xe1_nature_2014, li_triangular_2005} 
and mathematics \cite{droubay_palindromes_1995, zhi-xiong_properties_1994}.

%%%%%%%%%%%%%% FIGURE 1
\begin{figure}[t!]
\centering
\includegraphics[width=1\columnwidth]{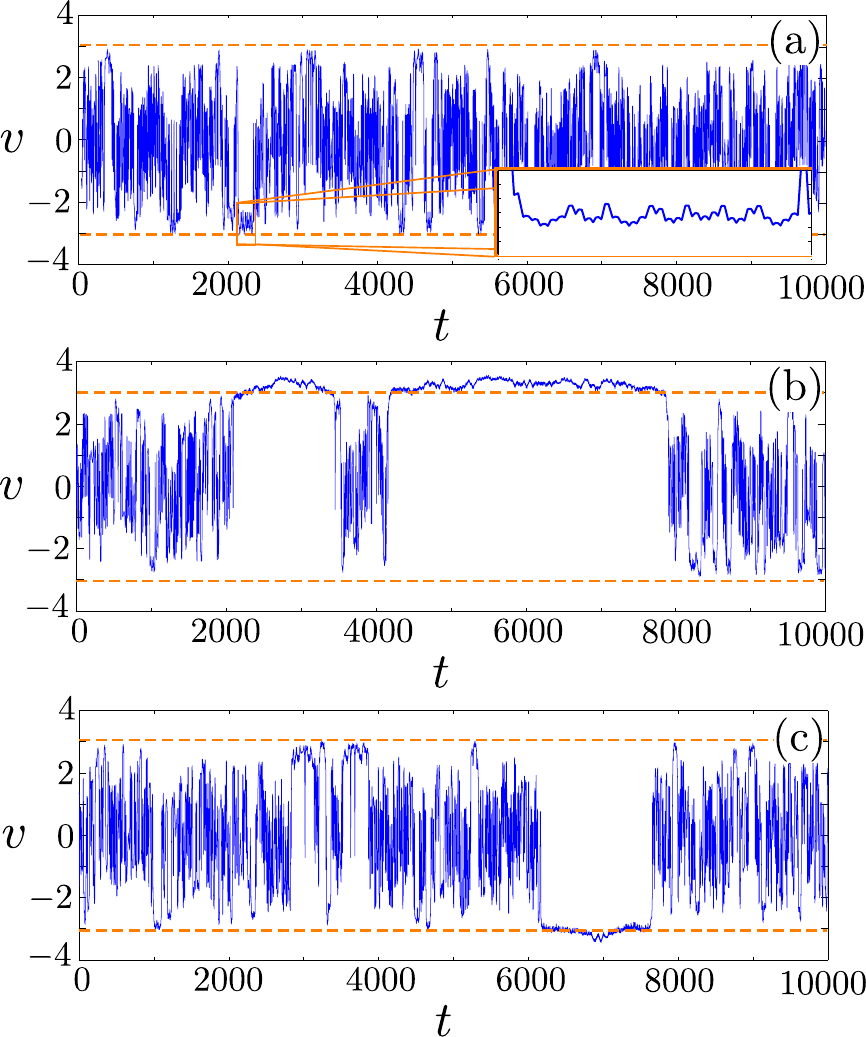}
\caption{\label{fig2} Exemplary trajectories $v(t)$ for the periodic (a), randomized (b) and the quasiperiodic lattice (c). In all three cases, the horizontal line denotes $\pm v^{\text{C}}_{\text{max}}$ (see Fig. \ref{fig1} (b)).
The inset in (a) shows a zoom into a typical stickiness event. Parameters are $V_\mathcal{A}=1.5$ (same as in Fig.\ref{fig1}(b)) and $V_\mathcal{B}=1.0$. Remaining parameters are the same as in Fig \ref{fig1}.} 
\end{figure}
%%%%%%%%%%%%%%%%%%%%%%%%%%%%%%%%%%%%%%%%%%%%%%%%%%%%%%%%%%%%%%%%%%%%%%%%%%%%%%%%%%%%%%%%%%%%%%%%%%%%%%%%%%%%%%%%%%%%%%%%%%%%%%%%%%%%%%%%%%%%%%%%%%%%%%%%%%%%%%%%%%%%%%%%%%%%%%%%%%%%%%%%%%%%%%%%%
\section{Motion in periodic, quasiperiodic and randomized driven lattices: Basic concepts.}
\label{sec2}
% In this section we will introduce some basic notions regarding the particle motion in driven lattice systems of the form of Eq. (\ref{Hamiltonian}).

In the periodic case, the setups mixed phase space can be visualized conveniently by the 
Poincar\'e surface of section (PSS). Here we denote 
velocities and phases $\phi\equiv (t \mod T)$
at positions $X^{\text{PSS}}= \{x,\ x = n\times L\}$ for $n \in \mathbb{N}$ (cf. Fig.\ref{fig1} (a)).
The for this manuscript relevant extract of the resulting PSS for a PL is shown in Fig.\ref{fig1} (b), revealing the typical ingredients: a 'chaotic sea', regular or 'ballistic' islands embedded in it and finally invariant curves confining the chaotic sea at higher velocities
(because of the time reversal symmetry
of the used Hamiltonian, the PSS is mirror symmetric around $v=0$).
An exemplary trajectory of the PL with initial conditions belonging to the chaotic sea is shown in Fig.\ref{fig2} (a) and shows a mostly erratic motion with frequent changes of magnitude and sign of the velocity,
accompanied by phases of motion where its velocity only fluctuates slightly; see e.g. inset of Fig.\ref{fig2} (a).
These 'stickiness' events are known to be quite generic for Hamiltonian systems and
simply put, originate from the fact that a chaotic trajectory gets drawn in by the intricate network of stable and 
unstable fixed points surrounding a regular structure which borders the chaotic sea \cite{lichtenberg_regular_1992}.
Furthermore, the maximal speed of a trajectory in the chaotic sea is denoted by $v^{\text{C}}_{\text{max}}$ 
(see horizontal dashed lines in Fig.\ref{fig1} (b) and Fig.\ref{fig2}).

For the randomized lattice, there is no such bound on the particles energy and it is in fact expected that the RL features Fermi acceleration as was demonstrated for comparable, randomized setups \cite{karlis_fermi_2007}. 
If we again consider an exemplary trajectory for the RL (Fig.\ref{fig2} (b)),
we see, in some analogy to the PL, an apparently irregular motion at velocities corresponding to the chaotic sea of the PL, which is interrupted by long unidirectional flights at higher velocities. A similar behavior can be observed for the sample trajectory in the quasiperiodic lattice (Fig.\ref{fig2} (c)), where again the particle motion at small velocities with $|v| \lesssim |v^{\text{C}}_{\text{max}}|$ is accompanied by large fluctuations in the velocity and is interrupted by long flights at higher velocities.
Hence, at least from this simple analysis based on sample trajectories, it appears that differences in the dynamical properties of the three studied lattice types are manifest mostly in the dynamics at 
$|v|\gtrsim |v^{\text{C}}_{\text{max}}|$ rather than in the low energy regime. 
%This is also rather intuitive as at low energies, the dynamics is mostly governed by the scattering properties of single barriers    
% The remainder of this manuscript is devoted to systematically investigate if and how any of the structural properties of the Fibonacci sequence translate into dynamical properties of the non equilibrium dynamics
% of particles. A natural approach to do so, is thereby to compare the dynamics in the FL with the corresponding dynamics in randomized and periodic lattices.
%Interestingly, at this point one might already argue that the motion during such a long flight event as shown in the inset of Fig. REF seems to resemble more the motion in the PL as in the RL. However, these are just sample trajectories

\section{Flight lengths in periodic, quasiperiodic and randomized driven lattices.}
\label{sec3}
% The remainder of this manuscript is devoted to the question if and how any of the structural properties of the Fibonacci sequence translate into dynamical properties of the non equilibrium dynamics
% of particles. A natural approach to analyze this, is to compare the dynamical evolution of particle ensembles in the FL with the corresponding dynamics in randomized and periodic lattices.

We now focus on a systematic investigation of the question if and how any of the structural properties of the Fibonacci sequence translate into dynamical properties of the nonequilibrium dynamics
of particles. As indicated above, a promising candidate for an effect where the periodic, randomized and quasiperiodic lattice significantly differ from one another are long flight events at velocities 
$|v| \gtrsim |v^{\text{C}}_{\text{max}}|$. 
Here, a particle traverses many barriers and thus correlations between lattice sites even on large scales can be expected to play a role. 
%While for the PL lattice, as explained above, the long flights are caused by the stickiness of the trajectory to a regular structure, their origin in the RL and particularly in the FL is not obvious.  

In order to investigate these long flight events quantitatively we calculate 
the flight length distribution $\Gamma(\Delta x)$ of particles, where 
the flight length $\Delta x$ is defined as the distance that a particle travels between two consecutive flips of the sign of its velocity. 
Particularly, for large $\Delta x$ the three different lattices types can be expected to deviate from one another, which we will explore in the following.
Numerically, $\Gamma(\Delta x)$ is obtained by 
propagating $N=2\times 10^4$ particles up to $t_{\text{max}}=10^8\times T$ with randomized initial velocities $-0.1<v_0<0.1$, so that all initial conditions would be located within the chaotic sea of the PL.
The starting positions are chosen randomly within the interval $5\times 10^8 + 10^3 < x_0/L < 5\times 10^8-10^3$ and  
numerical convergence with respect to $N$ and $t_{\text{max}}$, as well as independence from the choice of the initial positions was checked very carefully. 
%%%%%%%%%%%%%% FIGURE 3
\begin{figure}[t!]
\centering
\includegraphics[width=1\columnwidth]{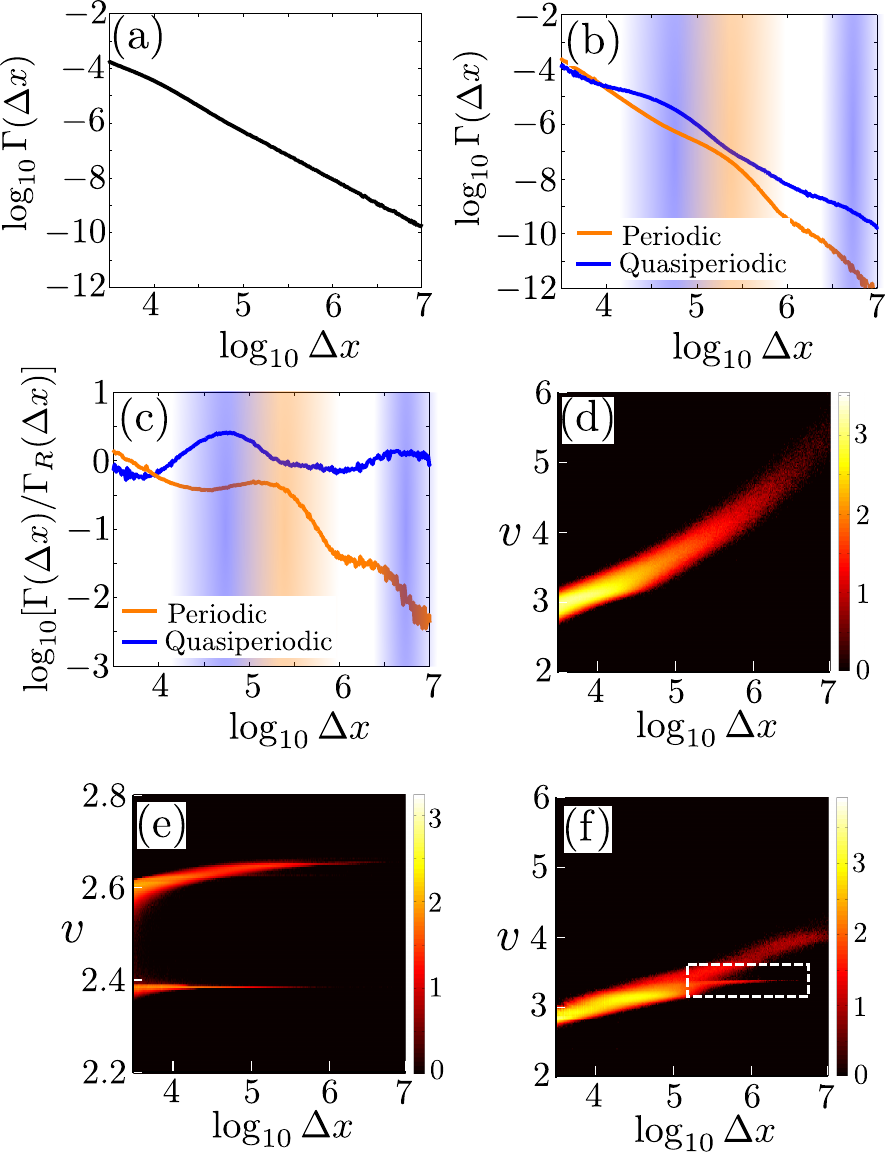}
\caption{\label{fig3} (a) Flight length distribution $\Gamma(\Delta x)$ for large $\Delta x$ in double logarithmic representation for the randomized lattice.
(b) $\Gamma(\Delta x)$ for periodic and quasiperiodic lattice. Shaded intervals correspond to extraordinarily long flights. (c) Ratio of $\Gamma(\Delta x)$ of the periodic or quasiperiodic 
lattices and $\Gamma(\Delta x)$ of the randomized lattice. Shaded intervals are identical as in (b).
Also shown are velocity resolved flight length distributions $\Gamma(\Delta x,\bar v)$ for the random (d), periodic (e) and 
quasiperiodic lattice (f). In the latter case, the dashed rectangle highlights the horizontal branch of long flights.} 
\end{figure}
%%%%%%%%%%%%%%%%%%%%%%%%%%%%%%%%%%%%%%%%%%%%%%%%%%
For the RL (Fig.\ref{fig3} (a)) we observe, 
to good approximation, for several orders of magnitude  a power law dependence $\Gamma(\Delta x)\propto (\Delta x)^{\alpha_{\text{RL}}}$ with some 
exponent $\alpha_{\text{RL}}$. 
%which is in fact the characteristic result for the tail of a Lévy flight length distribution (REF).
In some sense, this simple power law decay of $\Gamma(\Delta x)$ for the RL can be seen as a benchmark for the two other setups, as it represents the result for a completely uncorrelated lattice. Hence any deviations from 
$\Gamma(\Delta x)$ for the PL and particularly for the FL can be expected to relate to structural properties of the phase space of the corresponding lattice. 
In fact, the flight length distribution for the PL (Fig.\ref{fig3} (b)), 
while still featuring an overall polynomial decay, also reveals deviations from the pure power law like behaviour and shows a small amplitude oscillation around $5 \lesssim \log \Delta x \lesssim 5.5$. 
% Hence, in this window of lengths, $\Gamma(\Delta x)$ decays slower than one would expect from a power law fit to $\Gamma(\Delta x)$ for $\log \Delta x \lesssim 5$. 
Interestingly, these deviations from the power law like decay of $\Gamma(\Delta x)$ can also be observed for the FL. 
% Hence, again around certain values of $\Delta x$, the flight length distribution decays slower with increasing $\Delta x$. 
This can be seen even clearer when calculating the ratio of $\Gamma(\Delta x)$ for the PL or the FL w.r.t. the RL (see Fig.\ref{fig3} (c)). Here, in particular for the FL,
pronounced maxima 
are apparent at distinct flight lengths, indicating that certain $\Delta x$ are favored for the PL and the FL when comparing their flight length distributions to the one of the RL.

More insight into this effect can be obtained by calculating velocity resolved flight length distributions $\Gamma(\Delta x,\bar v)$ (see Figs.\ref{fig3} (d), (e) and (f)), where $\bar v$
is the average velocity
for a given flight of length $\Delta x$. 
(Similar to before, the time reversal symmetry of the Hamiltonian ensures that $\Gamma(\Delta x,\bar v)=\Gamma(\Delta x,-\bar v)$).
For the RL, we observe that  $\Gamma(\Delta x,\bar v)$ is concentrated around higher velocities, for longer flight lengths $\Delta x$. 
This is in accordance with the observation made from the sample trajectories, namely that as soon as $|v| < |v^{\text{C}}_{\text{max}}|$ the velocity sign changes rapidly.
Contrarily, once a particle reaches the high velocity regime, 
the particles kinetic energy is large compared to the lattice potential and the influence of the lattice potential on the particles velocity can expected to be small. Hence the particles velocity change
upon collision with a barrier
tends to be smaller the higher its velocity is, which supports the effect of longer unidirectional flights at higher velocities. 
% behavior can be understood intuitively by the following line of arguments:
% as seen above for the sample trajectories, particles performing long flights have typically large velocities ($|v| > |v^{\text{C}}_{\text{max}}|$), since as soon as $|v| < |v^{\text{C}}_{\text{max}}|$ the velocity sign changes rapidly.
% 
% Once a particle enters the high velocity regime however
% 
% Thus, the higher the velocity of a particle is, the longer it takes to reach the low energy regime again and consequently the particle may perform a longer flight. 
% Adding to this is the fact that, once a particle reaches the high velocity regime, i.e. a regime where its kinetic energy is large compared to the lattice potential, the influence of the lattice potential on the particles velocity can expected to be small. Hence the rate of the particles velocity change is smaller the higher its velocity is, which again supports the effect of longer unidirectional flights at higher velocities. 
Apparently, for the PL this simple line of arguments fails and $\Gamma(\Delta x,\bar v)$ looks qualitatively 
different (see Fig.\ref{fig3} (e)) revealing that $\Gamma(\Delta x,\bar v)\neq 0$ only along two branches centered around $\bar v \approx 2.4$ and $\bar v \approx 2.6$ respectively.
Firstly, as there is an upper bound of $v^{\text{C}}_{\text{max}}$ on the particles velocity in the PL, this bound holds of course also for the average velocities $\bar v$
and hence $\Gamma(\Delta x,\bar v)=0$ for $\bar v > v^{\text{C}}_{\text{max}} $ (keep in mind that the particle ensemble used to determine the flight length distributions is started with low energies, and is hence located entirely within the chaotic sea). Secondly, 
%$\Gamma(\Delta x,\bar v)$ is notably nonzero only within two narrow velocity intervals around $\bar v \approx 2.4$  
%and $\bar v \approx 2.6$. 
the reason for the appearance of these two branches can be understood conveniently by considering again the systems PSS (Fig.\ref{fig1} (b)). 
As mentioned above, long unidirectional flights in the PL are closely related to the stickiness of trajectories at regular structures which bound the chaotic sea. Apparently, there are two notable of such regular structures present: the 
chain of three islands around $v\approx 2.5$ as well as the first invariant spanning curve (FISC) acting as an upper bound of the chaotic sea at around 
$v\approx 2.9$.
Please note, that the three islands should indeed be interpreted as a single regular structure, since they share a common central orbit with a periodicity of three unit cells. 
Indeed one can check by inspecting the corresponding trajectories, that the flights at large $\Delta x$ around $\sim 2.4$ are caused by particles getting sticky to the island chain, while the branch around $\sim 2.6$ 
is caused by particles becoming sticky to the FISC.
The fact that both branches in 
$\Gamma(\Delta x,\bar v)$ appear to be at slightly smaller velocities than the associated regular structures has in fact a very simple explanation. For the PSS (Fig.\ref{fig1} (b)), the velocity of a particle 
is denoted at positions between scatterers, and hence at positions where the potential is zero. While passing through the lattice, the particle has to surpass the repulsive barriers, and thus momentarily 
its kinetic energy is lowered. Hence the average velocity of a particle moving along some regular structure can indeed be expected to be smaller then the velocity suggested by the PSS as shown in Fig.\ref{fig1} (b). 

Finally, lets turn our attention to the quasiperiodic case (Fig.\ref{fig3} (f)). Again, $\Gamma(\Delta x,\bar v)$ reveals the overall trend that longer flights possess larger 
average velocities, as already observed for the RL. Strikingly, we also see a horizontal branch centered around $\bar v \approx 3.2$, similarly to the two branches as observed for the PL. 
These, however, were remnants of regular structures of the PL's phase space, which -particularly in the case of regular islands- can be traced back to a 
synchronization of the particle motion with the lattice oscillation. As this, apparently, hinges on the periodicity of the lattice, it is now an intriguing 
question what the cause of the horizontal branch in $\Gamma(\Delta x,\bar v)$ of the FL is.

%%%%%%%%%%%%%%%%%%%%%%%%%%%%%
\section{Transient motion in quasiperiodic lattices.}
\label{sec4}

\begin{figure}[t!]
\centering
\includegraphics[width=1\columnwidth]{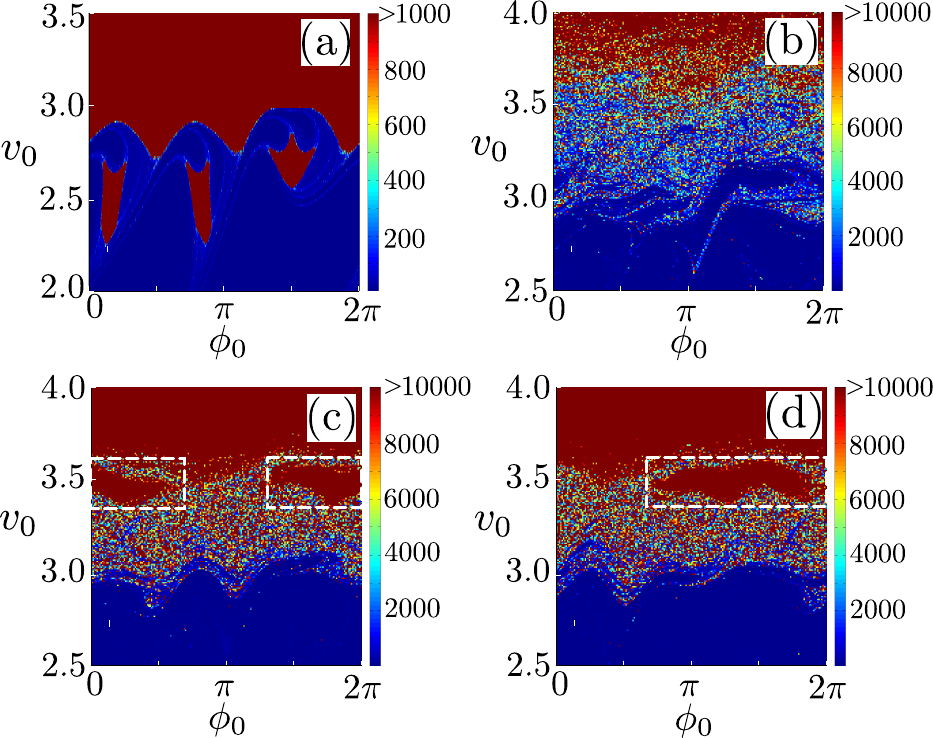}
\caption{\label{fig4} Flight length $\Delta x(x_0,\phi_0, v_0)$ starting at $x_0=0$ as a function of the initial phase and initial velocity for the periodic (a), randomized (b) and quasiperiodic (c) lattice.
(The shorter cutoff in the PL is for illustrative purpose only).
In (d) we show $\Delta x (x_0=100\times L,\phi_0, v_0)$ for the quasiperiodic case. Dashed rectangles highlight plateaus of extraordinarily long flights. Parameters are the same as in Fig.\ref{fig2}. } 
\end{figure}

While for the PL, all the regular structures of the corresponding phase space can be investigated conveniently by means of the PSS, the same procedure can not be applied to the FL, simply because 
the PSS inherently exploits the systems spatial periodicity. Hence, we must opt for a different approach and again turn to an observable related to the flights lengths. 
Here, we calculate the flight length of a given initial condition $(x_0,\phi_0,v_0)$ by propagating particles until their velocity passes $v=0$ for the first time.
At this point the modulus of the current positions $x$ of the particle minus $x_0$ gives the flight length $\Delta x (x_0,\phi_0,v_0)$ for this particular initial condition.
% Furthermore, since some initial conditions my lead to infinitely long flights we need to introduce a cutoff, which we set as $\Delta x_c = 10^4 \times L$ for the RL and the FL and as $\Delta x_c = 10^3 \times L$ for the PL
% (the shorter cutoff in the PL is for illustrative purpose only). 
The results for the PL, RL and the FL are shown in Figs.\ref{fig4} (a), (b) and (c) respectively for an initial position $x_0=0$.
For the PL, we very clearly see the counterparts of the regular structures as present in the PSS (cf. Fig.\ref{fig1} (b)). If the trajectory is started within one of these structures, it performs an 
unidirectional motion through the lattice 
and $\Delta x (x_0,\phi_0,v_0)$ is -in fact- infinite. 
%%%%%%%%%%%%%% FIGURE 4
%%%%%%%%%%%%%%%%%%%%%%%%%%%%%%%%%%%%%%%%%%%%%%
In this sense, calculating $\Delta x (x_0,\phi_0,v_0)$ can be seen simply as an alternative approach to determine the regular structures in the phase space of the PL. 
Its major advantage is that it does not intrinsically rely on the spatial periodicity of the system and can thus be applied for non periodic lattices also. Here, we, of course, have to keep in mind that
$\Delta x (x_0,\phi_0,v_0)$ is expected to depend on $x_0$ and in particular does not obey $\Delta x (x_0,\phi_0,v_0)=\Delta x (x_0 \pm n\times L ,\phi_0,v_0)$ as is does for the PL. Nevertheless, we will see that 
calculating the flight lengths for some exemplary $x_0$ for the RL and the FL does indeed reveal some valuable insight.
Starting with the RL (Fig.\ref{fig4} (b)), we observe no apparent separation between regular and diffusive motion as seen in the PL, and -in fact- it is reasonable to assume that every trajectory will eventually pass
$v=0$, thus leading to a finite flight length $\Delta x$ for all initial conditions. As a general trend, we again see that larger initial velocities $v_0$ tend to lead to longer flights which agrees well with the discussion concerning the 
flight length distribution $\Gamma(\Delta x,\bar v)$ (Fig.\ref{fig3} (d)). Finally, for the FL (Fig.\ref{fig4} (c)), $\Delta x (0,\phi_0,v_0)$ is qualitatively different from both the periodic and the randomized case. Here, we see a sharp 
increase of the flight length at around $v_0 \sim 3$. Furthermore, we see a plateau of extraordinarily long flights centered around $\phi_0 \sim 0 / 2\pi$ and $v_0 \sim 3.5$, which is very much reminiscent 
of the regular islands as seen for the PL. 
Interestingly, like the regular islands in the PL, the plateau falls together with a horizontal branch in the flight lengths distribution $\Gamma(\Delta x, \bar v)$ as shown in Fig.\ref{fig3}(f)
(for the same reason as before the plateaus' velocity in $\Delta x (0,\phi_0,v_0)$ appears slightly higher than the velocity of the corresponding branch in $\Gamma(\Delta x,\bar v)$).
% This similarity is underlined by the fact that, like the regular islands in the PL, the plateaus fall together with a horizontal branch in the flight length distribution 
% $\Gamma(\Delta x,\bar v)$ (Fig.\ref{fig3} (f)), where for the same reason as before the plateaus' velocity in $\Delta x (0,\phi_0,v_0)$ appears slightly higher than the velocity of the corresponding branch in $\Gamma(\Delta x,\bar v)$.  
At this point the question arises, how $\Delta x (x_0, \phi_0,v_0)$ changes upon changing $x_0$. Exemplarily, we show the result for $x_0=100\times L$ in Fig.\ref{fig4} (d), revealing that the observed plateau appears again within the same velocity interval but centered around a different phase.
By repeating this for various $x_0$ we can convince ourselves that the appearance of a plateau of long flights at velocities of $v_0 \sim 3.5$ seems to be a 'global' property of the FL, rather than a 
peculiarity for some distinct parts of the lattice.
Despite the similarities between the plateau as observed in the FL and the regular islands in the PL, there is also a major difference, which is the finite flight length within the FL 
even for trajectories started within the plateau, thus making this a phenomenon of a transient dynamics.

Lets us briefly summarize what we know about the motion in the FL so far:
Apparently, we have found domains of initial conditions leading to  
exceptionally long unidirectional flights. In particular, these defy the simple overall trend of longer flights for higher velocities
thereby contrasting the dynamics in the RL. Additionally, the flight lengths remain finite which -in turn- is in contrast to motion on a ballistic island of the PL. 
However, all the described features are reminiscent of a stickiness event of a trajectory in the PL. Hence, it appears as if trajectories in the FL would follow some 
phase space structure to which they become sticky for a long time, but are eventually able to escape. Naturally, the question arises what this phase space structure and its physical origin is and 
maybe more importantly if we can somehow deduce and understand is location around $v \sim 3.5$ (and $\phi \sim 0$ for $x_0=0$). 

%%%%%%%%%%%%%% FIGURE 5
\begin{figure}[t!]
\centering
\includegraphics[width=1\columnwidth]{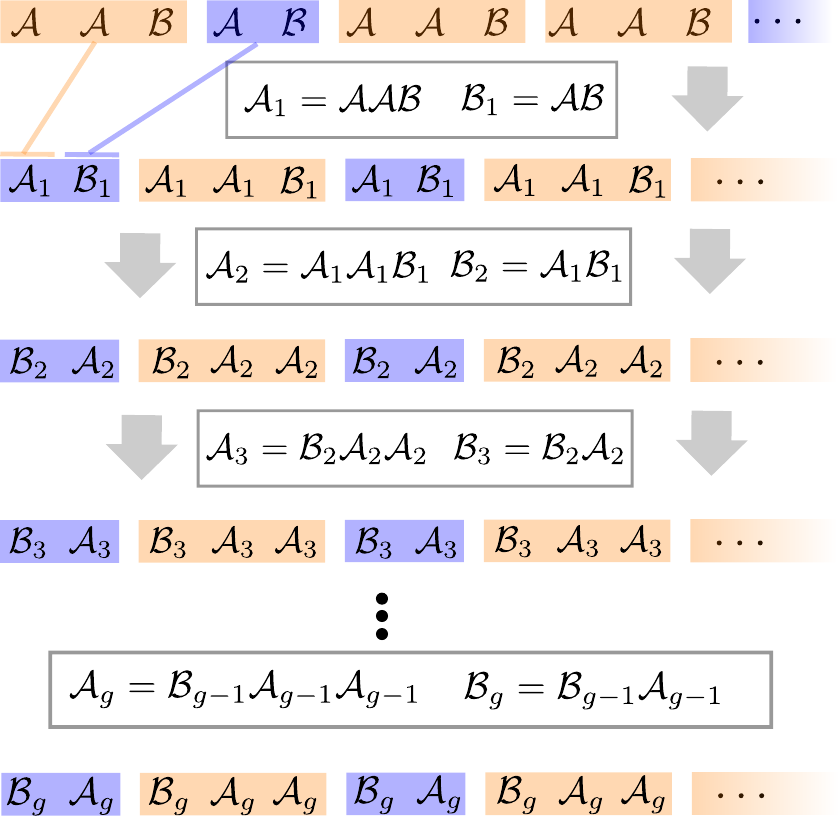}
\caption{\label{fig5} Block decomposition of the Fibonacci lattice in symbolic notation according to the decomposition rule given in Eqs. (6) and (7).
The first row depicts the first few elements of the Fibonacci sequence (cf. Eq.(\ref{sequence}) where a $1$ (0) corresponds to a barrier of type $\mathcal{A}$ ($\mathcal{B}$). At the same time, this first row
is the 'zeroth generation' of the decomposition hierarchy.
All further rows show block decompositions of increasing generations.}
\end{figure}

%%%%%%%%%%%%%%%%%%%%%%%%%%%%%
\section{Block decomposition of the Fibonacci Lattice.}
\label{sec5}

At this point, we need to make use of some distinct properties of the FL. In particular, we will argue that it can be decomposed into building blocks on various hierarchical levels.
Based on this block decomposition of the FL, we will construct a set of PSS of periodic lattices and their corresponding Poincar\'e Maps, which govern the dynamics in the FL on different length scales.
Finally, we show how invariant subsets of these Poincar\'e maps are related directly to the observed long ballistic flight events.

\subsection{Poincar\'e maps and their application to randomized systems.}
For the PL, 
we defined the Poincar\'e surfaces to be at positions $X^{\text{PSS}}= \{x,\ x = n\times L\}$ for $n \in \mathbb{N}$ (cf. Fig.\ref{fig1} (a)).
Subsequent coordinates on these surfaces of a trajectory moving through the lattice are then linked via the Poincar\'e map:   
\begin{equation}
(\phi_{k+1},v_{k+1})=\mathcal{M}_{\mathcal{A}}(\phi_k,v_k), 
\end{equation}
which is thus determined implicitly
by the scattering properties of the lattice barriers (the subscript '$\mathcal{A}$' denotes that, as before, the PL consists of $\mathcal{A}$-type barriers). 
Note that dynamical processes occurring on length scales below the distance of adjacent 
Poincar\'e surfaces are not resolved by $\mathcal{M}_{\mathcal{A}}$ for the given choice of the surfaces. For example  
orbits trapped between two positions of adjacent Poincar\'e surfaces which are present even for oscillating repulsive barriers (see \cite{koch_dynamical_2008}),
are not captured (but these are also not relevant for our work).
Main features of $\mathcal{M}_{\mathcal{A}}$ can be read off directly from the setups PSS (Fig.\ref{fig1} (b)). Particularly a stable fixed point $(\phi_f, v_f)$
of a given period $p$:   
\begin{equation}
 (\phi_f, v_f) = \mathcal{M}^p_A (\phi_f, v_f),
\end{equation}
where the superscript denotes a $p$-fold application of $\mathcal{M}_A$, is made apparent as $p$ regular islands in the PSS.
Thereby, each of the fixed point surrounding closed curves constitutes an invariant set under the action of $\mathcal{M}^p_A$.

Equivalently, we can describe the dynamics in nonperiodic lattices by means of successive applications of the 
Poincar\'e maps $\mathcal{M}_A$ and $\mathcal{M}_B$. 
% Let us now address the question how the notion of Poincar\'e Maps maybe applicable to nonperiodic, randomized lattices.
% First, we have to fix the positions of the Poincar\'e surfaces. 
% For starters, let us simply adopt the one from the periodic case, that is between any two adjacent scatterers. 
% As we allow for two different barrier types, $\mathcal{A}$ and $\mathcal{B}$, our setup inherits also two 
% different Poincar\'e maps: $\mathcal{M}_A$ and $\mathcal{M}_B$ and hence subsequent coordinates on the surfaces are given either by 
% $(\phi_1,v_1)=\mathcal{M}_{\mathcal{A}}(\phi_0,v_0)$ or by $(\phi_1,v_1)=\mathcal{M}_{\mathcal{B}}(\phi_0,v_0)$.
An intriguing question is now, whether 
a randomized lattice may allow for periodic motion on the level of Poincar\'e maps. 
One straightforward way how this could be realized, is by demanding that $\mathcal{M}_A$ and $\mathcal{M}_B$ share a common fixed point: 
$(\phi_f, v_f) = \mathcal{M}_A (\phi_f, v_f)= \mathcal{M}_B (\phi_f, v_f)$. 
If such a point exists, one might say that $(\phi_f, v_f)$ constitutes a fixed point of the dynamics in the entire nonperiodic lattice.
While by fine tuning of parameters it might indeed be accomplishable to match fixed points of $\mathcal{M}_A$ and $\mathcal{M}_B$, in a generic setting this cannot be expected to happen. 
Also,  
even if such a point exists, in order for it to be stable, the surrounding invariant sets of $\mathcal{M}_A$ and $\mathcal{M}_B$ would also have to be invariant under the action of both $\mathcal{M}_A$ and $\mathcal{M}_B$.
This seems to be even 
harder to accomplish by means of fine tuning parameters and in fact we see no such stable fixed points in both studied nonperiodic cases.
Finally, for nonperiodic lattices, 
fixed points of $\mathcal{M}_A$ or $\mathcal{M}_B$ of order $p>1$, corresponding to ballistic unbounded motion, are not relevant as these would require
a repeating sequence of $\mathcal{A}$ and $\mathcal{B}$ barriers.

\subsection{Symmetry adopted Poincare Maps of the Fibonacci lattice.}

Naively, one might argue that in order for the FL to support ballistic islands, again one would have to match 
fixed points of $\mathcal{M}_A$ and $\mathcal{M}_B$ and their invariant subsets. 
As we will see in the following, this is however too simplistic and the FL requires a more sophisticated analysis.
At this point, we would 
like to stress that we checked the validity of the numerical results, presented in the following, beyond the for this manuscript relevant scales.

The key idea is to realize that the FL can be decomposed into building blocks as illustrated in Fig.\ref{fig5}.  
In the first decomposition, we define the building blocks to be $\mathcal{A}_1\equiv \mathcal{A} \mathcal{A} \mathcal{B}$ and $\mathcal{B}_1\equiv \mathcal{A} \mathcal{B}$.
If we again focus on unbounded motion, we can now describe the dynamics on the level of the Poincar\'e maps $\mathcal{M}_{\mathcal{A}_1}$ and $\mathcal{M}_{\mathcal{B}_1}$, which iterate
trajectories between positions between adjacent blocks.
This decomposition can be continued, by defining the 'next generation' of blocks as: 
\begin{eqnarray}
\mathcal{A}_g= \mathcal{A}_{g-1} \mathcal{A}_{g-1} \mathcal{B}_{g-1}, \ \mathcal{B}_{g}= \mathcal{A}_{g-1} \mathcal{B}_{g-1} \ \text{for } g\leq2 \\
\mathcal{A}_g= \mathcal{B}_{g-1} \mathcal{A}_{g-1} \mathcal{A}_{g-1}, \ \mathcal{B}_{g}= \mathcal{B}_{g-1} \mathcal{A}_{g-1} \ \text{for } g>2 
\label{block decomposition}
\end{eqnarray}
with the corresponding Poincar\'e maps  $\mathcal{M}_{\mathcal{A}_g}$ and $\mathcal{M}_{\mathcal{B}_g}$ and with $\mathcal{A}_0 \equiv \mathcal{A}$
and $\mathcal{B}_0 \equiv \mathcal{B}$. 
Hence, the FL allows for unbounded regular motion, if two Poincar\'e Maps of any given generation feature two identical regular structures (in contrast to the RL, where only the two maps for $g=0$ are relevant). 
%%%%%%%%%%%%%% FIGURE 6
\begin{figure}[t!]
	\centering
	\includegraphics[width=1\columnwidth]{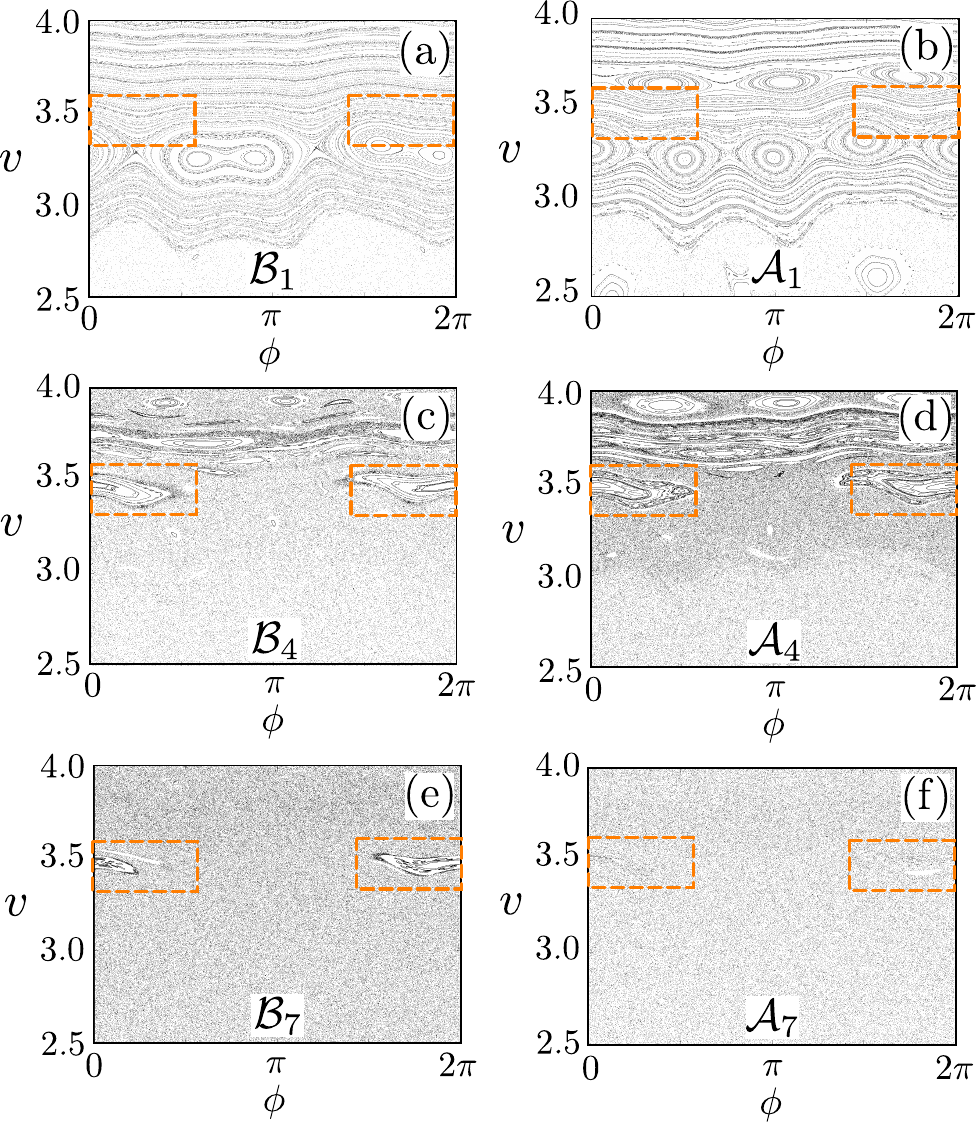}
	\caption{\label{fig6} Poincar\'e surfaces of section of lattices consisting of periodic repetitions of blocks: $\mathcal{B}_1$ (a), $\mathcal{A}_1$ (b),
		$\mathcal{B}_4$ (c), $\mathcal{A}_4$ (d), $\mathcal{B}_7$ (e) and $\mathcal{A}_7$ (f). Dashed rectangles are at the same locations as in Fig.\ref{fig4} (c) denoting a
		domain of long flights in the quasiperiodic lattice. Remaining parameters are as in Fig. \ref{fig2}.} 
\end{figure}

The invariant subsets of the Poincar\'e maps of any generation, can be visualized conveniently by means of the PSS of the corresponding periodic system (e.g. $\mathcal{A}_g\mathcal{A}_g\mathcal{A}_g...$)
with Poincar\'e surfaces between each adjacent building blocks. 
While, the 'zeroth generation' PSS corresponding to $\mathcal{A}_0$ 
is already shown in Fig.\ref{fig1} (b), 
some of the relevant PSS of various higher generations are shown in Fig.\ref{fig6}. We find that the PSS of blocks of some generations 
feature indeed a regular islands around the same phase space coordinates as the plateau of long ballistic flights as observed in the FL (cf. Fig.\ref{fig4} (c)).

In order to understand how these regular structures are linked with the long flights in the FL, 
lets consider the PSS corresponding to $\mathcal{A}_4$ and  $\mathcal{B}_4$ (Figs.\ref{fig6} (c) and (d)).
For example a trajectory starting at $x_0=0$ and with $(\phi_0, v_0)$ 
corresponding exactly to the fixed point centering the regular island of $\mathcal{M}_{\mathcal{B}_4}$ will pass the first surface of section at $x=34\times L$
(as this is the length of one $\mathcal{B}_4$ block) with the same coordinates $(\phi_1, v_1)=(\phi_0, v_0)$. The next block is of type $\mathcal{A}_4$ (and thus of length $55\times L$) and consequently,
the coordinates on the next Poincar\'e surface at $x=89\times L$ are given by  $(\phi_2, v_2)=\mathcal{M}_{\mathcal{A}4}(\phi_1, v_1)$. Even though $(\phi_1, v_1)$ does not correspond exactly to the fixed point of $\mathcal{M}_{\mathcal{A}_4}$, it does correspond to one of the invariant curves of the surrounding regular island and hence the trajectory will surpass the block confined to this particular invariant curve. Because the regular islands of both maps $\mathcal{M}_{\mathcal{A}_4}$
and $\mathcal{M}_{\mathcal{B}_4}$ are rather similar to one another, we can expect that the trajectory needs many such iterations before it can finally 
leave this particular domain of phase space, which ultimately causes the observed stickiness and equally the long ballistic flights at this particular velocity domain.

Even though one may describe the dynamics on different lengths scales in the FL by means of any of the possible decompositions, we see that 
only the PSSs of blocks of some particular generations contain notable common regular structures which then relate to domains of extraordinarily long ballistic flights.
We find numerically, that the appearance of similar regular structures in the PSS of $\mathcal{A}_g$ and $\mathcal{B}_g$ is suppressed more strongly with increasing $g$ 
with an increasing difference in the potential heights $V_{\mathcal{A}}$ and $V_{\mathcal{B}}$. In this way, one may -to some extent- choose which of the generations of the decomposition 
support long ballistic flights in the FL and thus also manipulate the length scale of these long flight events. For example, we find that by setting $V_{\mathcal{A}}=1.5$ as before and 
$V_{\mathcal{B}}=0.1$ (instead of $V_{\mathcal{B}}=1.0$) that the two maxima in the flight length distribution in the FL (as shown in Fig. \ref{fig3} (c)) are shifted by roughly one order of magnitude to smaller 
$\Delta x$ as compared to the case of $V_{\mathcal{B}}=1.0$ . Additionally, the regular structures in the hierarchical PSSs decay approximately one generation earlier, matching the observation of the 
shorter preferred length scale in the flight length distribution.

% Since, the description on the level of a certain decomposition naturally induces a length scale,
% Hence, we find that the certain phenomena can be traced back to the PSS of blocks of some specific generations, and thus of some specific length scale, which ultimately sets a highly non trivial length scale for dynamics in the FL.

\section{Conclusion}
\label{sec6}

We have investigated the chaotic dynamics of classical particles exposed to a periodically driven, spatially quasiperiodic lattice potential. As two points of references, we compare
our results to periodic- and fully randomized lattices and indeed find unique features of the particle dynamics for the quasiperiodic lattice. Specifically, we show that 
particles perform exceptionally long ballistic flights at distinct velocities. Since the usual tools as commonly applied in the investigation of periodic systems, such as 
Poincar\'e surfaces of sections, intrinsically rely on the spatial periodicity of the system, they cannot be applied straightforwardly here.
However, we show how a suitable set of Poincar\'e surfaces of periodic lattices provides the decisive insights into the dynamics of the quasiperiodic lattice. These Poincar\'e surfaces 
and their corresponding Poincar\'e maps are introduced naturally to the system by an underlying hierarchy of block decompositions of the lattice. Thereby, each Poincar\'e map associated to a decomposition
of a given level of the hierarchy describes the particle dynamics on a different length scale and we show how regular structures of these maps translate directly into the observed domains of 
long ballistic flights in the quasiperiodic lattice. Even though the block decompositions work up to arbitrarily large length scales, which of these scales are actually of relevance to the dynamics 
is determined by the scattering properties of the individual barriers constituting the lattice. Hence, the shown results are caused by an intricate interplay of the global structures of the quasiperiodic
lattice on the one hand, and of the 'local' scattering properties of individual barriers on the other hand.

\section*{Acknowledgments}

We thank B. Liebchen for many helpful discussions.

%%%%%%%%%%%%%%%%%%%%%%%%%%%%%%%%%%%%%%%

\bibliography{Bib_save}{}

%%%%%%%%%%%%%%%%%%%%%%%%%%%%%%%%%%%%%%%%%%%%%%%%%%%%%%%%%%%%%%%%%%%%%%%%%%%%%%%%%%%%%%%%%%%%%%%%%%%%%%%%%%%%%%%%%%%%%%%%%%%%%%%%%%%%%%%%%%%%%%%%%%%%%%%%%%%%%%%%%%%%%%%%%%%%%%%%%%%%%%%%%%%%%%%%%     

\end{document}